\documentclass[prl,twocolumn,showpacs,showkeys]{revtex4}

\usepackage{amssymb, amsmath, amsfonts, latexsym, verbatim, mathrsfs}

\usepackage{graphicx}

\begin{document}

\title{Entanglement magnification induced by local manipulations}

\author{Raffaele Romano}

\email{rromano@ts.infn.it} \affiliation{Department of Theoretical
Physics, University of Trieste, Italy}


\begin{abstract}

\noindent We study the entanglement capability of the evolution of a
pair of qubits evolving under unitary dynamics, when the local
dynamical parameters cannot be modified during the time-evolution.
Unlike the fast local control regime, we find that local and
non-local contributions to the dynamics are strictly interconnected.
Moreover, it is possible to strongly increase the entanglement
capability by suitably initializing the characteristic energies of
the two parties.

\end{abstract}

\pacs{03.67.Mn, 03.67.-a, 02.30.Yy}

\keywords{entanglement generation, local control}

\maketitle


{\it Introduction.---} A basic ingredient for the implementation of
quantum technologies is the ability to control some fundamental
processes involving a pair of qubits, that is, the capacity of
influencing the dynamics of this system through external actions.
When the interaction with the environment is negligible, the system
dynamics is given by the family of unitary transformations $U_t$
generated by the total Hamiltonian $H_T = H_1 + H_2 + H_I$, where
$H_1$, $H_2$ are one-qubit contributions, and $H_I$ represents the
interaction. This term is responsible for the generation of the
peculiar quantum correlations called {\it entanglement}. The growing
interest in the quantum theories of information and computation is
largely due to these correlations, and to their potential
applications. Therefore, it is of fundamental relevance to
characterize the entanglement, to understand its properties and to
study the processes leading to its creation. The relevant quantities
involved are determined by the specific physical realization of the
qubits pair.

The standard control setting is the so-called {\it Local Unitary
control (LU)}, where the control operations act locally (they
influence $H_1$ and $H_2$, not $H_I$). Moreover, motivated by the
available technologies, it is usually assumed that these operations
are performed instantaneously \cite{khan,benn} (e.g., by means of
short laser pulses). An arbitrary unitary transformation $U_t$ can
be written as
\begin{equation}\label{eq02}
    U_t = A_n \otimes B_n e^{-i H t_n} \cdots e^{-i H t_2} A_1 \otimes
    B_1 e^{-i H t_1} A_0 \otimes B_0,
\end{equation}
representing a succession of entangling evolutions, acting for time
intervals $t_1 \ldots t_n$, interspersed by instantaneous local
evolutions, represented by the $A_i$ and $B_i$ operators, that can
be arbitrarily manipulated. In general $H \ne H_I$, since the
physical interaction can be used to simulate the dynamics generated
by a different non-local contribution \cite{benn}. In (\ref{eq02}),
the non-local and local parts are mutually independent.

Other control schemes have been developed for unitary dynamics,
inspired by different experimental scenarios. In particular, the
results presented in this work are relevant for the {\it indirect
control} methods, in which an auxiliary system (ancilla) is used to
manipulate the target system through their mutual interaction
\cite{roma2,fu}.

In the framework of LU control, the processes involving the qubits
pair have been widely investigated in the past years. For fast local
controls, the time-optimal generation of entanglement has been
considered in \cite{dur}, using the idea of {\it entanglement
capability} of the interaction. For an arbitrary unitary operator,
the maximal achievable entanglement, and the corresponding
uncorrelated initial state, have been characterized in \cite{krau}.
A geometrical characterization of the entangling gates has been
given in \cite{zhan}, in terms of the coefficients describing the
entanglement production. Other relevant topics, not directly
connected to the entanglement generation, have been discussed in the
literature, as the simulation of non-local gates
\cite{benn,khan,vida,hamm,hase}, and the ability of transmitting
classical as well as quantum information \cite{hamm}.

Even if usually well justified, the expression (\ref{eq02}) is only
an approximation, since it involves instantaneous actions. In this
work, we relax this assumption by considering the opposite regime,
in which the controls cannot be modified during the time evolution.
In this case, it is not possible to distinguish between independent
local and non-local contributions, as in (\ref{eq02}). It turns out
that the entangling part of the dynamics inherits an explicit
dependence on $H_1$ and $H_2$ as long as $[H_1 + H_2, H_I] \ne 0$.
Local control and entangling dynamics become deeply interconnected,
and this dependence can be used to manipulate the non-local part by
means of local control. In this letter we derive the entanglement
capability in this regime, and show that particular choices of the
local parameters can highly increase it.

Our analysis complements previous results obtained under the fast LU
control assumption. Moreover, it is of interest in the context of
indirect control methods, where the entanglement between target and
auxiliary systems is fundamental to perform manipulations, and the
parameters of the ancilla have to be fixed accordingly.


{\it Cartan decomposition of unitary operators.---} We assume that
the time-evolution of the pair of qubits is given by a family of
unitary maps generated by the total Hamiltonian. A fundamental
property, known as Cartan decomposition (or canonical
decomposition), factorizes an arbitrary unitary operator $U \in
U(4)$ as
\begin{equation}\label{eq01}
    U = L A K,
\end{equation}
where $L, K \in SU(2) \otimes SU (2)$ are local operators, and $A =
e^a$, with $a$ an element of the Cartan subalgebra of the Lie
algebra $\mathfrak{u} (4) = span (i \sigma_j \otimes \sigma_j, j =
x, y, z)$, $\sigma_j$ the Pauli operators for the two subsystems.
The only operator that can correlate the two systems is $A$, that is
called the entangling part. $L$ is irrelevant when dealing with
entanglement generation.

This decomposition clarifies the role of the 16 parameters entering
an arbitrary unitary transformation $U (4)$: the local contributions
are characterized by twelve of them, the composite evolution by
three of them, and finally there is a not-relevant overall phase. In
particular,
\begin{equation}\label{eq01bis}
    a = -i \sum_j \theta_j \sigma_j \otimes \sigma_j
    = -i \sum_j \lambda_j \vert j \rangle \langle j \vert,
\end{equation}
where $\theta_j$ ($j = x, y, z$) and $\lambda_j$ ($j = 1, \ldots 4$)
are real constants embodying the entanglement capability of the
channel. In (\ref{eq01bis}), we have introduced the eigenvalues
$\lambda_j$ and eigenvectors $\vert j \rangle$ of $a$ (the so-called
magic basis \cite{hill}, given by the Bell states up to total
phases). We notice that the $\lambda_j$ are constrained by ${\rm
Tr}\, a = 0$, then they sum up to zero. It is always possible to
rearrange the coefficients as $\frac{\pi}{4} \geqslant \theta_x
\geqslant \theta_y \geqslant \theta_z \geqslant 0$, using their
properties of symmetry ($\theta_j \rightarrow \frac{\pi}{2} -
\theta_j$) and periodicity ($\theta_j \rightarrow \theta_j + k_j
\frac{\pi}{2}, k_j \in \mathbb{Z}$). The relation between the two
families of parameters is given by
\begin{eqnarray}\label{eq01tris}
    &\lambda_1 = \theta_x - \theta_y + \theta_z, \qquad
    &\lambda_2 = -\theta_x + \theta_y + \theta_z, \nonumber \\
    &\lambda_3 = -\theta_x - \theta_y - \theta_z, \quad \,
    &\lambda_4 = \theta_x + \theta_y - \theta_z.
\end{eqnarray}
The aforementioned entanglement capability of the interaction is
defined as $h = \theta_x + \theta_y$ (e.g. see \cite{dur}).

Notice that the decomposition (\ref{eq01}) justifies the expression
(\ref{eq02}); the non-local part is fully parameterized by three
real constants, independent of the local actions.


{\it Dynamical evolution.---} The most general Hamiltonian terms are
given by
\begin{eqnarray}\label{eq03}
    H_1 = \omega_1 \vec{n} \cdot \vec{\sigma} \otimes {\mathbb I}, \qquad H_2 = \omega_2
    {\mathbb I} \otimes \vec{m} \cdot \vec{\sigma}, \\ \nonumber H_I = \sum_{ij} c_{ij}
    \sigma_i \otimes \sigma_j, \qquad \qquad
\end{eqnarray}
where $i$ and $j$ range over $\{x, y, z \}$, $\omega_1$ and
$\omega_2$ are the characteristic energies of the two subsystems,
$\vec{n}$ and $\vec{m}$ are real unit vectors, $\vec{\sigma}$ is the
vector of Pauli matrices and the coefficients $c_{ij}$ form a real
matrix. Without loss of generality, we will consider representations
of the Pauli matrices, for the two subsystems, such that this matrix
is diagonal, $c_{ij} = c_i \delta_{ij}$.

The local actions consist of arbitrary preparations of $\vec{n}$,
$\vec{m}$, $\omega_1$, and $\omega_2$. In order to have manageable
expressions, we assume that only the characteristic energies can be
modified, and we fix $\vec{n} = \vec{m} = (0, 0, 1)$. We further
consider $\omega_i \geqslant 0$, $i = 1, 2$. The Cartan
decomposition of the unitary transformation $U_t$ acting on the
system is written as
\begin{equation}\label{eq05}
    U_t = e^{-i H_T t} = L_t A_t K_t,
\end{equation}
and the dependence on time is made apparent. We are interested in
the relevant contributions for the entanglement generation, that is
$A_t$ and $K_t$. It is possible to compute
\begin{equation}\label{eq06}
    U_t = \left[%
\begin{array}{cccc}
  e^{-i c_z t} \varphi_1(t) & 0 & 0 & e^{-i c_z t} \varphi_4(t) \\
  0 & e^{i c_z t} \varphi_2(t) & e^{i c_z t} \varphi_3(t) & 0 \\
  0 & e^{i c_z t} \varphi_3(t) & e^{i c_z t} \varphi_2^*(t) & 0 \\
  e^{-i c_z t} \varphi_4(t) & 0 & 0 & e^{-i c_z t} \varphi_1^*(t) \\
\end{array}%
\right]
\end{equation}
where
\begin{eqnarray}\label{eq07}
    \varphi_1(t) &=& \cos{\Omega_1 t} - i \, \Bigl( \frac{\omega_1 +
    \omega_2}{\Omega_1} \Bigr) \sin{\Omega_1 t}, \nonumber \\
    \varphi_2(t) &=& \cos{\Omega_2 t} - i \, \Bigl( \frac{\omega_1 -
    \omega_2}{\Omega_2} \Bigr) \sin{\Omega_2 t}, \nonumber \\
    \varphi_3(t) &=& - i \, \Bigl( \frac{c_x + c_y}{\Omega_2} \Bigr) \sin{\Omega_2 t},
    \nonumber \\
    \varphi_4(t) &=& - i \, \Bigl( \frac{c_x - c_y}{\Omega_1} \Bigr) \sin{\Omega_1 t},
\end{eqnarray}
with frequencies
\begin{eqnarray}\label{eq08}
    \Omega_1 &=& \sqrt{(c_x - c_y)^2 + (\omega_1 + \omega_2)^2},
    \nonumber \\
    \Omega_2 &=& \sqrt{(c_x + c_y)^2 + (\omega_1 - \omega_2)^2}.
\end{eqnarray}

In order to find the terms of the decomposition (\ref{eq05}), it is
convenient to represent all the operators in the magic basis, in
which the local contributions become orthogonal matrices
$\tilde{L}_t$ and $\tilde{K}_t$, and the non-local part
$\tilde{A}_t$ is diagonal,
\begin{equation}\label{eq05bis}
    \tilde{U}_t = \tilde{L}_t \tilde{A}_t \tilde{K}_t.
\end{equation}
Since $\tilde{U}_t^T \tilde{U}_t = \tilde{K}_t^T \tilde{A}_t^2
\tilde{K}_t$, it is possible to determine $\tilde{A}_t$ and
$\tilde{K}_t$ by considering the eigenvalues and eigenvectors of
this operator. Applying the same procedure to $\tilde{U}_t
\tilde{U}_t^T$ it is possible to find also $\tilde{L}_t$, however
this contribution is not relevant for the purposes of this paper.
Denoting the eigensystem as
\begin{equation}\label{eqes}
    \tilde{U}_t^T \tilde{U}_t \vec{v}_i (t) = \varepsilon^2_i (t)
    \vec{v}_i (t), \quad i = 1, \ldots, 4,
\end{equation}
we obtain
\begin{eqnarray}\label{eq09}
    \varepsilon^2_{1,2} (t) &=& e^{2 i c_z t} \Bigl( \chi_1 (t) \pm i \, \sqrt{1 - \chi_1^2 (t)}
    \Bigr), \nonumber \\
    \varepsilon^2_{3,4} (t) &=& e^{- 2 i c_z t} \Bigl( \chi_2 (t) \pm i \, \sqrt{1 - \chi_2^2 (t)}
    \Bigr),
\end{eqnarray}
where we have defined two real-valued functions,
\begin{eqnarray}\label{eq10}
    \chi_1 (t) &=& 1 - 2 \, \Bigl( \frac{c_x + c_y}{\Omega_2} \Bigr)^2
    \sin^2{\Omega_2 t}, \nonumber \\
    \chi_2 (t) &=& 1 - 2 \, \Bigl( \frac{c_x - c_y}{\Omega_1} \Bigr)^2
    \sin^2{\Omega_1 t}.
\end{eqnarray}
Since $-1 \leqslant \chi_i (t) \leqslant 1$ ($i = 1,2$), it is
possible to check that $\vert \varepsilon_i (t) \vert = 1$, $i = 1,
\ldots 4$. The corresponding (non-normalized) eigenvectors are given
by
\begin{eqnarray}\label{eq10bis}
    \vec{v}_{1,2} (t) &=& (\mu_1 (t) \mp \nu_1 (t), 0, 0, 1)^T,
    \nonumber \\
    \vec{v}_{3,4} (t) &=& (\mu_2 (t) \mp \nu_2 (t), 0, 0, 1)^T,
\end{eqnarray}
where
\begin{equation}\label{eq10tris}
    \mu_1 (t) = \frac{\Omega_2}{\omega_1 - \omega_2} \cot{\Omega_2
    t}, \quad
    \mu_2 (t) = \frac{\Omega_1}{\omega_1 + \omega_2} \cot{\Omega_1
    t},
\end{equation}
and $\nu_i (t) = \sqrt{1 + \mu_i^2 (t)}$, for $i = 1, 2$. The
normalized eigenvectors form the rows of the matrix $\tilde{K}_t$.

\noindent The non-local part in the magic basis is given by
\begin{equation}\label{eq12}
    \tilde{A}_t = diag \bigl( \varepsilon_i (t), i = 1, \ldots 4
    \bigr),
\end{equation}
and, following equation (\ref{eq01bis}), it is possible to write
\begin{equation}\label{eq11}
    \varepsilon_j (t) = e^{- i \lambda_j (t)}, \quad j = 1, \ldots
    4.
\end{equation}
Therefore the Cartan coefficients are given by
\begin{eqnarray}\label{eq11bis}
    \lambda_{1,2} (t) &=& c_z t \pm \frac{1}{2} \arccos{\chi_2 (t)},
    \nonumber \\
    \lambda_{3,4} (t) &=& -c_z t \mp \frac{1}{2} \arccos{\chi_1 (t)},
\end{eqnarray}
and the expressions for the $\theta_i (t)$, $i = x, y, z$, can be
obtained inverting (\ref{eq01tris}). These parameters characterize
at every time the non-local contribution to the dynamics. Unlike the
arbitrarily fast LU control case, they are time-dependent real
functions that contain an explicit dependence on the local controls
through the parameters $\Omega_1$ and $\Omega_2$, unless $[H_1 +
H_2, H_I] = 0$. This condition is satisfied only in two cases:
either $\omega_1 = \omega_2 = 0$, or $c_x = c_y = 0$. In both cases
$\lambda_j (t)$ are linear in $t$ and independent of the local
controls, since the local dynamics and the interaction become
independent processes.

Finally, the relevant contributions can be rewritten in the original
basis. The local term can be cast in the form
\begin{equation}\label{eq13}
    K_t = \left[%
\begin{array}{cccc}
  0 & 0 & 0 & e^{-i \eta_2 (t)} \\
  0 & 0 & e^{i \eta_1 (t)} & 0 \\
  0 & e^{-i \eta_1 (t)} & 0 & 0 \\
  e^{i \eta_2 (t)} & 0 & 0 & 0 \\
\end{array}%
\right]
\end{equation}
where
\begin{equation}\label{eq13bis}
    \eta_i (t) = \sqrt{\frac{\nu_i (t) + \mu_i (t)}{\nu_i (t) - \mu_i (t)}},
    \qquad i = 1, 2.
\end{equation}
The non-local contribution has the form
\begin{widetext}
\begin{equation}\label{eq14}
    A_t = \frac{1}{2} \left[%
\begin{array}{cccc}
  \varepsilon_2 (t) + \varepsilon_3 (t) & 0 & 0 & \varepsilon_2 (t) - \varepsilon_3 (t) \\
  0 &  \varepsilon_1 (t) + \varepsilon_4 (t) &  \varepsilon_1 (t) - \varepsilon_4 (t) & 0 \\
  0 &  \varepsilon_1 (t) - \varepsilon_4 (t) &  \varepsilon_1 (t) + \varepsilon_4 (t) & 0 \\
  \varepsilon_2 (t) - \varepsilon_3 (t) & 0 & 0 &  \varepsilon_2 (t) + \varepsilon_3 (t) \\
\end{array}%
\right].
\end{equation}
\end{widetext}

\begin{figure}[t]
\begin{center} 
  \includegraphics[width=6.5cm]{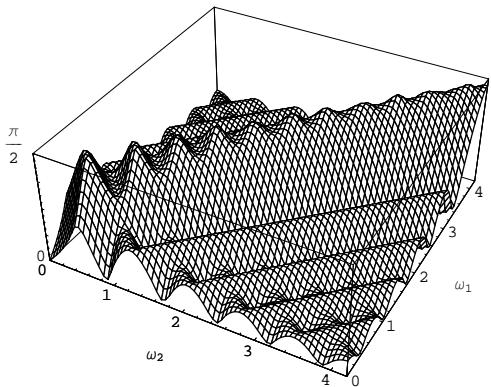} \\
   \caption{\footnotesize Dependence on $\omega_1$ and $\omega_2$ of the
   entanglement capability $h (t)$, for the Ising interaction $c_x = 1$,
   $c_y = c_z = 0$, and $t = \frac{3}{2} \pi$. The main peaks of $h (t)$
   correspond to $\omega_1 = \omega_2$. The other peaks are related to
   the extremal points of $\chi_i (t)$ with respect to $\omega_1$ and
   $\omega_2$, and $t$ fixed.}\label{fig1}
\end{center}
\end{figure}

{\it Entanglement capability and optimal input states.---} We are
now able to derive the entanglement capability $h(t)$ of $U_t$ and
study its properties. Considering (\ref{eq11bis}), it is possible to
obtain
\begin{equation}\label{eq15}
    h (t) = \frac{\pi}{2} \, k (t) + \frac{1}{2} \, s (t) \arccos{\chi_{i(t)} (t)},
\end{equation}
where $k (t)$, $s (t)$ and $i (t)$ are piecewise continuous
functions with values in $\mathbb{Z}$, $\{ -1, 1 \}$, and $\{ 1, 2
\}$ respectively, such that the hierarchy relations among the
$\theta_j (t)$ are fulfilled at every time $t$. The specific form of
these functions is not relevant for the purposes of this work. The
entanglement capability is a continuous function in $t$, with $0
\leqslant h (t) \leqslant \frac{\pi}{2}$, and it depends on the
local parameters $\omega_1$ and $\omega_2$ through the functions
$\chi_i (t)$, $i = 1, 2$. The extremal points for $\chi_i (t)$ can
be found from $\partial_t \chi_i (t) =
\partial_{\omega_1} \chi_i (t) =
\partial_{\omega_2} \chi_i (t) = 0$, whose relevant solutions are
\begin{equation}\label{eq16}
    \omega_1 \mp \omega_2 = 0, \qquad t = k \, \frac{\pi}{2 \, (c_x \pm
    c_y)},
\end{equation}
where the upper sign holds for $i = 1$, the lower for $i = 2$, and
$k \in \mathbb{Z}$. A typical dependence of $h (t)$ on $\omega_1$
and $\omega_2$ for a critical value of $t$ is represented in Fig.
\ref{fig1}. The complicate peak-valley pattern is determined by the
extremal points of the functions $\chi_i (t)$, according to
(\ref{eq15}). From (\ref{eq16}), it can be seen that the initial
preparation of the local parameters has a strong impact on the
evolution of the entanglement capability. In general, a sudden
change is observed when $\omega_1$ and $\omega_2$ match. This is
consistent with the behavior of the purification process discussed
in \cite{roma}. In fact, the entangling capability of the evolution
is fundamental in indirect control schemes.

An initial state $\vert \psi \rangle_0$ is transformed by the
evolution in the usually entangled state $\vert \psi \rangle_t = U_t
\vert \psi \rangle_0$. The maximal attainable entanglement depends
on the entanglement capability as well as on the initial state. It
is possible to characterize the set of the optimal input states by
solving the equation
\begin{equation}\label{eq17}
    \vert \psi \rangle_0 = K_t A_t^{\dagger} \vert \Psi \rangle,
\end{equation}
where $\vert \Psi \rangle$ is an arbitrary maximally entangled
state, $K_t$ and $A_t$ are expressed in (\ref{eq13}) and
(\ref{eq14}) respectively, and $K_t^{\dagger} = K_t$.


{\it Conclusions.---} When a two-qubits system is manipulated via
local controls, it is usually assumed that these actions can be
performed in an arbitrarily small time. Under this hypothesis, there
is a clear separation between local and non-local contributions in
the dynamics. The entangling capability of the evolution is a
constant embodying the interaction content of the dynamics, and
every initial factor state can produce the maximal amount of
entanglement, since it can be instantaneously transformed, by fast
local actions, in the optimal input state for the entanglement
generation.

In this paper we have explored the entanglement generation in the
opposite situation, where the local actions are fixed during the
evolution. We have found that, in this regime, the local and
non-local parts of the dynamics are strictly interconnected.
Considering a particular control model, we have found the expression
of the entanglement capability on the evolution time $t$, and on the
local parameters $\omega_1$, $\omega_2$. In particular, $h(t)$
reaches its global maxima periodically in $t$, under the condition
$\omega_1 = \omega_2$. Only some selected input states maximize the
entanglement production.

\begin{figure}[t]
\begin{center} 
  \includegraphics[width=6.5cm]{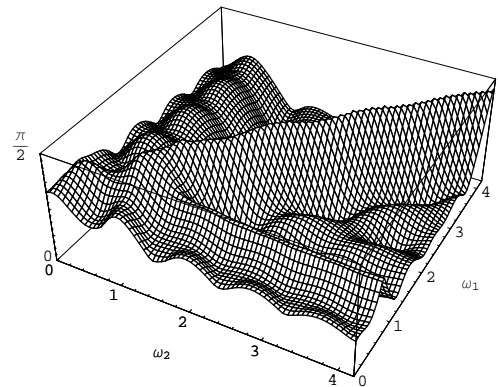} \\
   \caption{\footnotesize Dependence on $\omega_1$ and $\omega_2$ of the
   entanglement capability $h (t)$, with $c_x = -0.2$, $c_y = 0.5$, $c_z =
   -0.8$, $t = 3.7$,
   and $\vec{n}$ and $\vec{m}$ are the unit vectors corresponding to $(1, -3, 2)$
   and $(-2, 1, 4)$ respectively. The condition $\omega_1 = \omega_2$
   still plays a special role.}\label{fig2}
\end{center}
\end{figure}

If less restrictive control models are adopted, in general it is not
possible to obtain simple analytical expressions for the
entanglement capability. However, there is strong evidence that the
main features of the behavior of $h (t)$, described in this work, do
not depend on the particular choice of $\vec{n}$ and $\vec{m}$. In
fact, we have numerically computed $h (t)$ for a large sample of
Hamiltonian operators, with randomly distributed values of $c_x$,
$c_y$, $c_z$, $\vec{n}$ and $\vec{m}$. We have always found results
that are consistent with the analysis presented in this work, in
particular the insurgence of the peak of $h (t)$ when the two
characteristic energies match. An prototypical example of these
simulations is presented in Fig. \ref{fig2}, that exhibits the
behavior of $h (t)$ for a particular evolution with $\vec{n} \ne
\vec{m}$. The increase of $h (t)$ in correspondence of $\omega_1 =
\omega_2$ is not an artifact of the choice of the Hamiltonian terms
assumed in this work, it is rather a general phenomenon in the
two-qubits system.

The author acknowledges support from the European grant
ERG:044941-STOCH-EQ. Work in part supported by Istituto Nazionale di
Fisica Nucleare, Sezione di Trieste, Italy.

\end{document}